\documentclass[sigconf]{acmart}

\renewcommand\footnotetextcopyrightpermission[1]{} % removes footnote with conference information in first column

\setcopyright{none}

\acmConference[]{}{}{}

\AtBeginDocument{
    
}

\begin{document}

%% allowing the author to define a "short title" to be used on the page 
%% headers.
\title{Integration Barriers in Open-Source SSI Frameworks: An Exploratory Developer Experience Probe}

\author{Breno Cerqueira Reis Nakamura}
\affiliation{
  \institution{Instituto de Ciência e Tecnologia, Universidade Federal de São Paulo (UNIFESP)}
  \city{São José dos Campos}
  \state{SP}
  \country{Brazil}
}
\email{breno.nakamura@unifesp.br}

\author{Arlindo F. da Conceição}
\affiliation{
  \institution{Instituto de Ciência e Tecnologia, Universidade Federal de São Paulo (UNIFESP)}
  \city{São José dos Campos}
  \state{SP}
  \country{Brazil}
}
\email{arlindo.conceicao@unifesp.br}

%% By default, the full list of authors will be used on the page
%% headers. This list is often too long and will overlap
%% other information printed in the page headers. 
%% This command allows the author to define a more concise list
\renewcommand{\shortauthors}{Nakamura and Conceição}
\renewcommand{\shorttitle}{Integration Barriers in SSI Frameworks}

%% The abstract is a short summary of the work the paper presents.
\begin{abstract}
Self-Sovereign Identity (SSI) promises to decentralize digital identity, but widespread adoption remains hindered by integration complexity and tooling immaturity. This paper investigates the Developer Experience (DX) of open-source SSI tooling through an exploratory probe study. Nine developers with prior knowledge of decentralized identity concepts, representing early integrators building SSI applications, attempted core credential lifecycle tasks using Walt.id, Traction, and MetaMask. Our goal was to surface recurring integration barriers through qualitative thematic analysis of open-ended developer reports, complemented by task-level difficulty ratings. Our findings reveal a critical abstraction gap: while passive operations like credential receipt are relatively mature, active construction tasks, particularly schema customization, expose significant architectural friction. We identify that these barriers stem from inadequate API abstractions, brittle environment configurations, and documentation that fails to track the ecosystem's rapid evolution. These issues reflect structural design decisions in current frameworks. This study characterizes the structural integration barriers in the current SSI open-source ecosystem. To address the identified gaps, we propose three architectural shifts for developer tooling: web-based sandboxes, AI-assisted schema generators, and executable documentation strategies.
\end{abstract}

%% the presented work. Separate the keywords with commas.
\keywords{Self-Sovereign Identity, Developer Experience, Integration Barriers, Open-Source Frameworks, API Abstraction}

%% This command processes the author, affiliation, and title
%% "frenchspacing" avoids an additional space after a period at the end of a sentence.
\frenchspacing
\maketitle

\section{Introduction}

    Digital identity management is a cornerstone of the modern online experience~\cite{WorldEconomic}. As technology advances, digital wallets have evolved from simple storage tools to agents capable of managing digital credentials. In this context, Self-Sovereign Identity (SSI) represents the next evolutionary step~\cite{InevitableRise}, allowing individuals to maintain full control over their data rather than depending on centralized providers. 
    
    SSI is being explored in domains such as healthcare data sharing, academic credential verification, and decentralized finance~\cite{SurveySSI}. Large-scale initiatives such as the European Blockchain Services Infrastructure (EBSI)~\cite{ebsi} indicate that SSI is moving toward real-world deployment. However, this transition exposes an architectural gap: the open-source frameworks required to build SSI-based applications impose significant integration burdens on developers that the current literature has largely overlooked~\cite{Korir2022, Zaeem2021, SurveySSI}.

    Research on SSI has concentrated on two poles: the cryptographic and architectural design of the protocols, and the end-user experience of consuming credentials through mobile wallets~\cite{Korir2022, Zaeem2021}. Adjacent work in the Web3 domain has documented how poor developer tooling acts as a primary barrier to adoption, often linking inadequate interfaces directly to security failures~\cite{eskandari2018, voskobojnikov2021}. Yet, the Developer Experience (DX) of integrators (developers responsible for deploying SSI infrastructure and embedding credential workflows into software systems) remains undercharacterized. These actors occupy a critical position in the adoption chain: before any end-user can receive or present a Verifiable Credential, a developer must have successfully configured an issuer agent, defined a credential schema, and established the trust infrastructure. 

    This paper addresses this gap through an exploratory probe study. Nine developers with prior knowledge of SSI concepts (representing the early integrator profile) attempted to complete core credential lifecycle tasks using three open-source frameworks: Walt.id~\cite{Waltid}, Traction~\cite{Traction}, and MetaMask~\cite{Metamask}. To ensure technical rigor, the evaluation is grounded in the established functional requirements for decentralized identity~\cite{Li2025}. Our goal is to surface integration barriers through a qualitative analysis of developer-reported friction, complemented by task-level difficulty ratings. We hypothesize that if this technically proficient cohort encounters consistent failures, then the barriers are architectural and will therefore persist until the underlying tooling design changes.

Our contributions are:

\begin{itemize}
    \item An exploratory empirical probe characterizing integration barriers encountered by developers performing core SSI credential lifecycle tasks with open-source frameworks.
    \item A structured identification of recurring failure modes, including environment configuration fragility, inadequate API abstractions for schema engineering, and documentation that does not track rapid ecosystem changes.
    \item Architectural implications for the SSI open-source ecosystem, highlighting practical tooling needs such as web-based sandboxes, AI-assisted schema generators, and executable documentation to reduce integration friction.
\end{itemize}

The remainder of this paper is structured as follows. Section~\ref{section:background} introduces the foundations of SSI and frames integration quality as a software engineering concern. Section~\ref{section:relatedWork} surveys the literature and isolates the developer abstraction gap. Section~\ref{section:StudyDesign} describes the probe methodology, while Section~\ref{section:Results} presents the quantitative and qualitative findings. Section~\ref{section:threats} discusses threats to validity, and Section~\ref{section:conclusion} concludes with architectural directions and future work.

\section{Background}
\label{section:background}
This section introduces the core technical components of SSI and describes the role of digital wallets as primary integration agents. It then examines integration quality, including API design, documentation quality, and abstraction level, as a key software engineering factor for deploying effective SSI solutions.

\subsection{SSI Architecture and Core Concepts}

Self-Sovereign Identity (SSI) is a paradigm designed to give users full control over their digital identity data, shifting from traditional centralized or federated identity models toward decentralized ones~\cite{InevitableRise}. The SSI ecosystem is structured around three primary roles: the \textit{Holder}, who possesses and controls their identity data; the \textit{Issuer}, a trusted entity that creates and cryptographically signs credentials attesting to certain information; and the \textit{Verifier}, who requests and validates proofs from the Holder.

This trust triangle is built upon two W3C-standardized technical pillars: Decentralized Identifiers (DIDs)~\cite{did-w3c} and Verifiable Credentials (VCs)~\cite{vc-model}. A DID is a globally unique identifier that resolves to a DID Document containing public keys and service endpoints, enabling cryptographic verification without a central authority. A VC is the digital equivalent of a physical document (such as a driver's license or a diploma), cryptographically signed by its Issuer and structured according to a schema registered in a Reliable Data Registry~\cite{SurveySSI}. 

Each of these components (DID resolution, VC schema definition, cryptographic signing, and proof generation) represents a distinct API surface that a developer must interact with. Consequently, each layer carries its own configuration and dependency requirements, increasing the architectural complexity of SSI systems.

\subsection{Digital Wallets as Integration Points}

In SSI, the digital wallet functions as the primary software agent for the Holder, responsible for managing the user's DIDs and cryptographic keys, storing VCs received from different Issuers, and generating proofs for presentation to Verifiers~\cite{WhatIsIdentityWallet}. Because SSI systems depend on interoperability between different platforms, wallets also act as critical integration points across the ecosystem~\cite{InevitableRise, SurveySSI, why-needs}. Industry specifications, such as the W3C Verifiable Credentials Data Model v2.0~\cite{vc-model}, provide the normative framework for this cross-platform credential exchange.

The wallet is not merely a user interface but a complex integration point. It exposes the control layer to both end-users and the external systems that must issue, present or verify credentials. Open-source wallet frameworks extend this complexity further: a developer integrating an open-source SSI stack must configure agent communication, typically via DIDComm protocols, manage schema registries, orchestrate container-based deployments, and consume multiple API layers simultaneously. The quality of this integration experience, encompassing API clarity, documentation fidelity, and environment stability, determines whether the framework can be adopted in practice~\cite{Wallet-reference}.

\subsection{Integration Quality Requirements and Developer Experience}
\label{subsec:integration-quality}

Developer Experience (DX) is  an attribute that describes the friction developers face when
integrating a system, encompassing API learnability, documentation
accuracy, and the cognitive load imposed by the system's architectural
abstractions~\cite{NEW_Fagerholm_2012_DX, quality-metrics}.

Poorly abstracted APIs impose an \textit{abstraction gap}, a mismatch between the developer's expectations and the actual requirements of the API, leading to predictable integration errors~\cite{myers2016improving, NEW_Piccioni_2013_API}. In rapidly evolving ecosystems, outdated documentation further compounds this problem by making it harder for developers to understand system behavior and correct API usage~\cite{NEW_Robillard_2011_Docs}.

In the SSI context, achieving architectural properties such as \textit{control} and \textit{portability}~\cite{Cucko2022} requires developers to implement the functional requirements of credential creation, verification, and recovery~\cite{Li2025}. If the tooling exposes raw cryptographic primitives or depends on fragile container orchestration instead of providing clean abstractions, these properties remain strictly theoretical.

\section{Related work}
\label{section:relatedWork}

This section surveys literature on developer tooling in Web3 ecosystems and end-user experiences in SSI to identify the specific abstraction gap developers face when actively integrating decentralized frameworks.

\subsection{Developer Tooling Challenges in Web3 Ecosystems}
\label{subsec:web3-tooling}
Because SSI wallets share foundational components with cryptocurrency wallets (public-key cryptography, seed phrases, and irreversible transactions), the Web3 tooling literature provides a useful baseline for understanding integration friction in decentralized systems~\cite{voskobojnikov2021}. Studies in this area show that poor tooling quality is a primary barrier to adoption, independent of the quality of the underlying protocol, with abstraction failure identified as the root cause~\cite{eskandari2018}: tools expose the raw complexity of the cryptographic layer instead of providing task-appropriate interfaces.

At the integration layer, open-source frameworks in rapidly evolving ecosystems often introduce API changes that require significant adaptation effort from client developers~\cite{hora2018developers}. In the SSI ecosystem, where standards such as W3C VCDM are still evolving, developers following older tutorials may encounter APIs that differ substantially from documented examples, turning documentation maintenance into a critical part of the integration process itself.

\subsection{End-User Studies in SSI: A Contrasting Perspective}
\label{subsec:enduser-ssi}
The existing literature on SSI wallets has focused predominantly on end-user experience. Heuristic evaluations of commercial solutions such as Evernym and uPort found that documentation was frequently missing, too technical, or hard to locate~\cite{Zaeem2021}. Studies on lay users reveal that they struggle to understand where their data is stored, often conflating sovereignty with simple local storage~\cite{Korir2022}. Other studies applied Cognitive Walkthrough methods to common SSI tasks using binary pass/fail metrics that measure completion rates but miss the friction experienced along the way~\cite{abylay}.

Despite their contributions, these studies share a common scope: they evaluate the experience of \textit{consuming} credentials through finalized products, not the experience of \textit{building} the infrastructure that makes credential exchange possible. While this body of work establishes that even polished SSI products present experience challenges for end users, it does not characterize the integration barriers faced by developers working at the framework and tooling level.

\subsection{The Developer Gap: From Passive Usage to Active Construction}

The distinction between \textit{passive usage} (receiving and presenting credentials) and \textit{active construction} (deploying agents, defining schemas, configuring trust registries) is central to understanding the SSI adoption problem. A developer integrating an open-source SSI framework must simultaneously act as Issuer, Verifier, and infrastructure operator, with each role introducing different APIs and configuration requirements. In this multi-role scenario, a single issue, such as a misconfigured DID resolver or a broken container network, can affect multiple subsystems, blocking downstream tasks and making debugging more difficult~\cite{NEW_Fagerholm_2012_DX, myers2016improving}.

To the best of our knowledge, no prior empirical study has specifically investigated these integration challenges in open-source SSI frameworks. This study addresses this gap by evaluating the specific friction developers face during active framework integration.

\section{Study Design}
\label{section:StudyDesign}

This study is designed as an exploratory probe: a structured but small-scale empirical investigation intended to identify integration barriers in open-source SSI frameworks rather than produce statistically generalizable results. This approach is well established in software engineering research, where intentionally scoped studies with technically informed participants are recognized as an effective method for identifying failure modes and generating hypotheses for larger follow-up investigations~\cite{stol2018abc}.

\subsection{Study Context and Tool Selection}
\label{subsec:tools}

The study was conducted as part of an undergraduate elective course on Digital Identities at the Federal University of São Paulo (UNIFESP). Tool selection was left to participants to better approximate a realistic adoption scenario, where developers choose frameworks based on available documentation and community resources. Three frameworks were selected.

\textbf{Walt.id}~\cite{Waltid}, chosen by six participants, is an open-source decentralized identity infrastructure that exposes APIs, SDKs, and command-line tooling for wallet and credential management. Its prevalence among participants may indicate greater accessibility within the developer community.

\textbf{MetaMask}~\cite{Metamask}, selected by two participants, is a non-custodial Web3 wallet designed for Ethereum-based asset management and interaction with decentralized applications (dApps). Although it is a general-purpose Web3 tool rather than a native SSI framework, its inclusion served as a baseline for distinguishing SSI-specific usability challenges from general blockchain interaction issues.

\textbf{Traction}~\cite{Traction}, selected by one participant, is an enterprise-oriented tenant agent built on Hyperledger Aries Cloud Agent Python (ACA-Py), supporting credential exchange through DIDComm protocols. Its architecture targets organizational deployments and reflects a different design approach compared to Walt.id.

The execution environment reflected typical developer setups: personal computers running Linux or Windows, with Docker required for local agent deployment. Because container orchestration is a core integration requirement for all three frameworks, it was evaluated as a measurable component of the ``Environment Setup'' task rather than treated as a separate logistical precondition.

\subsection{Task Design}
\label{subsec:tasks}

Five tasks were defined to cover the complete credential lifecycle. These tasks reflect the functional requirements for decentralized identity applications established in the software engineering literature~\cite{Li2025}. This grounding ensures that the barriers identified in this study are relevant to production integration scenarios, not artifacts of an artificial laboratory setup. The essential use cases identified in the field were translated into specific developer tasks:

\begin{enumerate}
    \item \textbf{Environment Setup (Installation):} Assesses the complexity of initial framework configuration, including dependency management (Docker, network setup, environment variables) required to bring a functional SSI agent online.
    
    \item \textbf{Credential Issuance:} Evaluates the Issuer role: creating, signing, and transmitting a Verifiable Credential to a Holder.
    
    \item \textbf{Credential Receipt:} Evaluates the Holder role: receiving, parsing, and storing a credential issued by a third-party.
    
     \item \textbf{Verification:} Evaluates the Verifier role: requesting a credential presentation and cryptographically validating the proof against the Issuer's public key.
    
    \item \textbf{Schema Engineering (Customization):} The most demanding task, assessing the ability to define or modify credential schemas for specific use cases. This task acts as a stress test of the framework's abstraction layer, requiring the developer to operate at the level of JSON-LD structures and schema registries without high-level tooling support.
\end{enumerate}

The progression from Installation to Schema Engineering was designed to gradually expose developers to different layers of the SSI stack, allowing the study to identify where integration and abstraction difficulties became more significant.

\subsection{Participants and Data Collection}
\label{subsec:participants}

The study involved nine participants, all final-year undergraduate students with prior theoretical knowledge of SSI concepts. This profile approximates the \textit{early integrator}: a developer who understands the conceptual model of DIDs and VCs but has not yet accumulated hands-on experience with a specific framework. Early integrators are an important audience for open-source SSI tooling and are likely to play a central role in broader ecosystem adoption.

The sample size is consistent with the goals of an exploratory probe. In software engineering, small but technically homogeneous samples are commonly used in barrier discovery studies, where the objective is to identify recurring failure modes and understand their underlying causes rather than to measure their prevalence across a population~\cite{stol2018abc}. If technically informed participants encounter similar failures, this suggests that the barriers may be related to the tooling or architecture rather than to individual experience, and warrant further investigation regardless of sample size.

The experimental procedure followed a linear execution path. Participants attempted the five core integration tasks sequentially using their selected framework. To capture realistic developer troubleshooting behavior, no artificial time limits were imposed, allowing participants to consult documentation, debug configurations, and explore the tools naturally. Immediately following the practical attempts, data was collected via a structured questionnaire divided into three main sections: 

\begin{itemize}
    \item \textbf{Part 1: Initial Information.} Began by obtaining informed consent for academic data use, in compliance with data protection regulations. Participants then identified the specific SSI frameworks they had configured.
    \item \textbf{Part 2: Task Evaluation.} Formed the core of the instrument. A standardized evaluation block was applied sequentially to each of the five core tasks (Installation, Issuance, Receipt, Verification, and Customization).
    \item \textbf{Part 3: Final Feedback.} Concluded with open-ended questions capturing overarching difficulties not previously mentioned and perspectives on future uses for SSI.
\end{itemize}

Figure~\ref{fig:questionnaire} illustrates the core evaluation block (Part 2) applied to each task. While the instrument utilized standard terminology for task ease (ranging from 1 = Unacceptable to 5 = Excellent), our analysis operationalizes this metric as a diagnostic indicator of \textit{integration friction}. In this engineering context, an ``Unacceptable'' rating indicates a blocking architectural barrier preventing task completion, while ``Excellent'' represents seamless integration without framework resistance. The quantitative rating was immediately paired with an open-ended prompt to capture the specific technical blockers encountered (e.g., error messages, missing abstractions, or configuration failures).

\begin{figure}[htbp]
    \centering
    \fbox{
        \begin{minipage}{0.9\columnwidth}
            \vspace{0.2cm}
            {\small
            \textbf{Part 2: Task Ease Assessment} \\
            \textit{Current Task: [e.g., Schema Engineering]} \vspace{0.3cm}

            \textbf{1. Ease of [Current Task]:} \\
            \vspace{0.1cm}
            $\bigcirc$ 1 (Unacceptable) \quad
            $\bigcirc$ 2 (Poor) \quad
            $\bigcirc$ 3 (Fair) \\
            $\bigcirc$ 4 (Good) \quad
            $\bigcirc$ 5 (Excellent) \vspace{0.4cm}

            \textbf{2. What difficulties were faced during [Current Task]?} \\
            \vspace{0.1cm}
            \framebox[\linewidth]{\rule{0pt}{1.2cm} \color{gray}\textit{(Open-ended text area...)}}
            \vspace{0.2cm}
            }
        \end{minipage}
    }
    \caption{Visual representation of the core evaluation block applied to each task, combining a quantitative task rating with an open-ended prompt for technical blockers.}
    \label{fig:questionnaire}
\end{figure}

The open-ended responses were analyzed using thematic analysis. Participant comments were reviewed iteratively to identify recurring integration blockers, which were then grouped into broader structural categories representing systematic failure patterns in the developer workflow. Three dominant themes emerged from this process and are reported in Section~\ref{section:Results}.

Participants were informed that their responses would remain confidential, be used only for academic purposes, and appear under anonymized identifiers in published materials. The data collection instrument and anonymized dataset are publicly available, as described in the \textit{Artifact Availability} section.

\section{Results and Discussion}
\label{section:Results}

The results are presented in two parts. First, the aggregate friction scores across tasks reveal where the integration difficulties increase. Second, the thematic analysis of open-ended responses helps identify the underlying causes associated with those difficulties.

\subsection{Task Friction Analysis: The Abstraction Gap}
\label{subsection:frictionAnalysis}

\begin{figure}[ht]
    \centering
    \includegraphics[width=1\columnwidth]{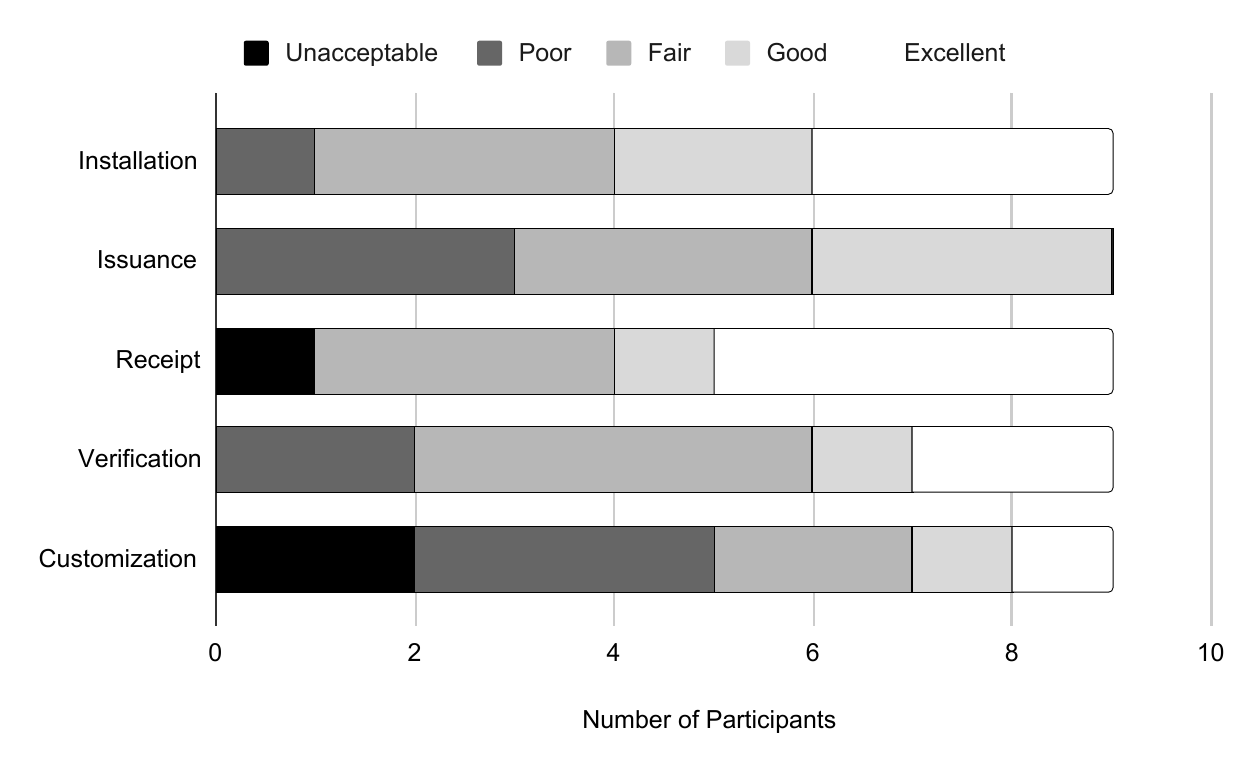}
    \caption{Distribution of developer-reported task evaluation scores using horizontal stacked bars. Scores range from 1 (Unacceptable, black) to 5 (Excellent, white). The shift from Receipt to Customization illustrates the abstraction gap: the point at which passive credential handling gives way to active schema construction.}
    \label{fig:task-score}
\end{figure}

Table~\ref{tab:resultados_quantitativos} presents the distribution of developer-reported friction scores across the five credential lifecycle tasks. Figure~\ref{fig:task-score} visualizes the full score distribution per task using horizontal stacked bar.

\begin{table}[ht]
\centering
\caption{Developer-reported task evaluation scores (1 = Unacceptable, 5 = Excellent).}
\label{tab:resultados_quantitativos}
\resizebox{\columnwidth}{!}{
\begin{tabular}{lcccc}
\hline
\textbf{Task} & \textbf{Mean} & \textbf{Median} & \textbf{Std. Dev.} & \textbf{Mode} \\
\hline
Installation   & 3.8 & 4.0 & 1.0 & 3 \& 5 \\
Issuance       & 3.0 & 3.0 & 0.8 & 3 \\
Receipt        & 3.8 & 4.0 & 1.3 & 5 \\
Verification   & 3.3 & 3.0 & 1.1 & 3 \\
Customization  & 2.6 & 2.0 & 1.3 & 2 \\
\hline
\end{tabular}
}
\end{table}

The data indicates a clear distinction between tasks involving passive credential handling and those requiring active framework construction. \textit{Credential Receipt} returned the highest aggregate scores (Mean: 3.8; Mode: 5), confirming that the Holder-side flow is relatively mature in current open-source tooling. \textit{Schema Customization} sits at the opposite extreme: Mean 2.6, Mode 2, with over half of participants reporting scores of 1 or 2.

This difference is substantial. Schema Customization corresponds directly to the \textit{Control} architectural requirement~\cite{Cucko2022}: the ability for a developer to define custom credential templates for a specific deployment context. The sharp reduction in scores for this task suggests that current frameworks become substantially harder to use as integration complexity increases particularly during tasks involving customization and deployment configuration. The high standard deviation (1.3) also suggests that outcomes varied sharply depending on which framework the participant used, a pattern explored further in the comparative analysis.

\subsection{Analysis of Reported Integration Barriers}
\label{subsection:thematicAnalysis}

The analysis of the open-ended responses revealed three recurring integration challenges. Each was associated with a different aspect of the SSI integration process.

\begin{enumerate}
    \item \textbf{Environment Fragility (Installation).} The aggregate friction score for Installation (Mean: 3.8) masks the specific difficulties encountered with dedicated SSI infrastructures. Participants configuring Walt.id reported consistent friction around container networking. The dependence on Docker and specific port configurations was a recurring hurdle. 
    \begin{quote}
    \textit{``It was necessary to discover that a recurring installation failure was due to attempting to use connection ports that were already in use.''} (Participant 1, Walt.id)
    \end{quote}
    For a technology aiming for sovereignty, the dependency on heavy runtime environments creates a high barrier to entry for self-hosting.

    \item \textbf{Documentation Deficits (Issuance):}
    Barriers in Credential Issuance were centered on information retrieval. Participants reported ``missing or outdated documentation'' and a ``trial-and-error process'' to find correct JSON templates. Furthermore, the documentation often covers only the ideal scenario. This directly corroborates previous findings in commercial apps~\cite{Zaeem2021}, suggesting that the documentation deficit is endemic to the SSI ecosystem, spanning from open-source SDKs to end-user products. In rapidly evolving ecosystems, documentation that is not continuously synchronized with the codebase degrades into a source of misinformation rather than a learning resource~\cite{hora2018developers}.

    \item \textbf{Abstraction Failure (Customization):}
    The most consequential theme emerged from the Customization task (Mean 2.6). Participants reported an inability to proceed due to ``severe tool limitations'' and the absence of examples:
    \begin{quote}
        \textit{``The process of issuing custom VCs is not documented and does not work as it should... There is no simple way to issue a custom credential.''} (Participant 7)
    \end{quote}
    This pattern is consistent with what Spolsky described as the \textit{Law of Leaky Abstractions}~\cite{spolsky2002law}: the principle that all non-trivial abstractions eventually expose the complexity of the layers they were designed to hide. In the SSI context, developers attempting to define a custom schema are forced to operate at the level of JSON-LD contexts, W3C VC Data Model types, and cryptographic binding specifications, layers that higher-level SDK abstractions would ideally reduce developers' exposure to. As a result, the functional requirement of \textit{Credential Creation}~\cite{Li2025} in a custom deployment context becomes difficult to achieve with current tooling, not because the standard itself lacks support for it, but because existing frameworks provide limited abstraction over the underlying complexity.
\end{enumerate}

\subsection{Comparative Analysis by Tool}
\label{subsection:ComparativeAnalysis}

Given the uneven distribution of tool selection (Walt.id: $n=6$, MetaMask: $n=2$, Traction: $n=1$), comparisons between frameworks should be interpreted as qualitative indicators of friction patterns rather than statistically conclusive rankings. Table~\ref{tab:raw_results} presents the mean task scores for each framework, while Table~\ref{tab:tool-comparison} summarizes the main integration blockers reported for each tool and the corresponding layer of the SSI stack involved.

\begin{table}[ht]
\centering
\caption{Mean task ease ratings (1--5) reported by participants for each framework.}
\label{tab:raw_results}
\resizebox{\columnwidth}{!}{
\begin{tabular}{lccccc}
\hline
\textbf{Framework ($n$)} & \textbf{Install.} & \textbf{Issue} & \textbf{Receive} & \textbf{Verify} & \textbf{Custom.} \\
\hline
Walt.id ($n=6$)       & 3.3 & 2.8 & 3.5 & 3.3 & 1.8 \\
MetaMask ($n=2$)      & 4.5 & 3.0 & 4.0 & 3.5 & 3.5 \\
Traction ($n=1$)      & 5.0 & 4.0 & 5.0 & 3.0 & 5.0 \\
\hline
\textbf{Overall Mean} & \textbf{3.8} & \textbf{3.0} & \textbf{3.8} & \textbf{3.3} & \textbf{2.6} \\
\hline
\end{tabular}
}
\end{table}

\begin{table}[ht]
\centering
\caption{Main integration blockers and affected layers for each framework.}
\label{tab:tool-comparison}
\begin{tabular}{p{1.5cm}p{3.2cm}p{2.5cm}}
\hline
\textbf{Framework} & \textbf{Main Blocker} & \textbf{Affected Layer} \\
\hline
Walt.id & Lack of abstraction for custom schema definition; undocumented API changes & SDK / Schema layer \\
MetaMask & Protocol mismatch between Web3 asset model and VC data model & Protocol layer \\
Traction & Webhook configuration and backend connection management & System integration layer \\
\hline
\end{tabular}
\end{table}

The disaggregated results suggest that the three frameworks present integration difficulties at different layers of the SSI stack. For Walt.id, the main challenges were concentrated at the SDK and schema abstraction layer. In particular, the low Customization score (1.8) was associated with the absence of any high-level schema builder and to undocumented API changes that invalidated participant workflows mid-task.

MetaMask presented a different pattern. Its reported difficulties were mainly related to the mismatch between a general-purpose Web3 asset wallet and the VC data model, rather than to SSI framework limitations. This distinction is relevant because it suggests that the abstraction issues observed in Walt.id are properties of dedicated SSI frameworks, not of decentralized tooling in general.

Traction provided an important contrast. It received high scores in Installation, Receipt, and Customization, the tasks where Walt.id failed. However, the participant reported significant difficulty with webhook configuration and backend service connectivity. This suggests that Traction reduces some schema-level integration barriers while introducing additional complexity at the system integration layer. Rather than eliminating integration friction entirely, the framework appears to shift part of the implementation burden to deployment and backend configuration tasks.

Taken together, these findings suggest that current open-source SSI frameworks still present significant integration challenges across different stages of the credential lifecycle. The reported difficulties do not appear to stem only from isolated implementation issues, but also from architectural and design choices that expose substantial underlying complexity to developers during integration.

\section{Threats to Validity}
\label{section:threats}

\textbf{Sample size and scope.} The probe involved nine participants, which precludes statistical generalization across the developer population. This is a deliberate property of the study design, not an oversight. In exploratory studies focused on barrier discovery, the objective is to identify recurring failure modes rather than establish prevalence rates~\cite{stol2018abc}. Larger follow-up studies may investigate the prevalence of these barriers across broader developer populations; the contribution of this study is the identification and characterization of the integration challenges themselves.

\textbf{Participant profile.} Participants were final-year undergraduate students with theoretical knowledge of SSI concepts but no prior hands-on experience with the evaluated frameworks. This profile approximates early integrators: developers familiar with the conceptual foundations of DIDs and VCs who are attempting to build their first SSI-based applications. As a result, the findings may underestimate the integration difficulties that less prepared developers would encounter in practice.

\textbf{Ecosystem volatility.} Walt.id and Traction are under active development, and the API behaviors, error messages, and documentation gaps reported in this study reflect the state of these frameworks at the time of data collection. Some of the specific issues identified may be resolved in future releases. However, the broader categories of integration difficulty observed in the study, such as abstraction and integration complexity challenges, may continue to affect developers as the SSI ecosystem evolves.

\textbf{Tool selection bias.} Participant-driven tool selection produced an uneven distribution across frameworks (Walt.id: $n=6$, MetaMask: $n=2$, Traction: $n=1$). As a result, comparisons between tools should be interpreted as qualitative observations rather than robust comparative evidence. Future studies may adopt a more balanced allocation across frameworks to better distinguish protocol-related difficulties from tool-specific implementation issues.

\section{Conclusion and Future Work}
\label{section:conclusion}

This study contributes an empirical characterization of the integration challenges faced by developers using open-source SSI frameworks. Through an exploratory probe involving nine early integrators performing core credential lifecycle tasks, we observed that integration difficulties became substantially more pronounced during active construction tasks, particularly schema engineering (Mean: 2.6 compared to 3.8 for Receipt tasks).

The qualitative findings suggest that many of these difficulties are related to architectural and abstraction challenges within current SSI frameworks. Participants frequently encountered situations in which low-level concepts, including JSON-LD context definitions, DID resolution mechanisms, and credential schema configuration, were exposed directly during integration tasks~\cite{spolsky2002law}. These issues became especially visible during custom credential creation tasks~\cite{Li2025}, where developers were required to coordinate multiple interdependent components simultaneously. Environment instability and documentation problems further increased the integration burden, suggesting that current SSI tooling still requires substantial architectural knowledge from developers during deployment and customization activities.

These findings have implications for SSI adoption. The \textit{Control} architectural requirement~\cite{Cucko2022}, the property that defines SSI as sovereign rather than merely decentralized, is only achievable if a developer can successfully define and deploy custom credential schemas. In our study, these customization tasks consistently produced the highest levels of difficulty among participants.

First, the high friction of local environment setup discourages initial exploration. The dependency on local Docker orchestration could be mitigated by web-based playgrounds or sandbox environments, allowing developers to explore Issuer-Holder-Verifier workflows without extensive local configuration. Second, schema engineering currently requires developers to interact directly with complex JSON-LD structures. This friction could be mitigated by AI-assisted tooling that translates natural language requirements into compliant credential templates, shielding developers from the underlying syntactic and cryptographic complexity. Finally, static documentation alone appears insufficient in rapidly evolving ecosystems. Interactive and executable documentation approaches, closely synchronized with the codebase, may provide more reliable support for developers as SSI frameworks continue to evolve.

Specifically, the development and empirical validation of AI-assisted schema generation tooling for SSI represents a direct next step, targeting the abstraction gap identified in the Customization task. Beyond tooling, longitudinal studies could investigate whether developer experience improves as W3C VCDM 2.0 stabilizes and SSI frameworks mature. Comparative studies involving low-code SSI platforms may also help determine whether higher-level abstractions reduce integration difficulties or simply shift them to other layers of the stack. Finally, future research could explore integration quality heuristics tailored to decentralized systems, addressing challenges related to key management and multi-component integration workflows in Web3 and SSI ecosystems.

Without accessible abstractions, accurate documentation, and stable deployment environments, building practical SSI applications will remain a significant bottleneck, limiting the transition of decentralized identity from a theoretical model into widely deployed infrastructure.

\section*{Artifact Availability}
The complete replication package for this study is publicly available on GitHub at \url{https://github.com/brenonak/ssi-integration-barriers}. The dataset includes the assessment instrument, raw task scores, and qualitative feedback.

\section*{Acknowledgements}
This work was supported by the São Paulo Research Foundation (FAPESP), grants \#2023/00783-7 and \#2025/06172-5.

\bibliographystyle{ACM-Reference-Format}
\bibliography{sample-base}

\end{document}